\documentclass[sigconf]{acmart}

\renewcommand\footnotetextcopyrightpermission[1]{}

\AtBeginDocument{%
  \providecommand\BibTeX{{%
    \normalfont B\kern-0.5em{\scshape i\kern-0.25em b}\kern-0.8em\TeX}}}

\usepackage[utf8]{inputenc} 
\usepackage[T1]{fontenc}    
\usepackage{hyperref}       
\usepackage{url}            
\usepackage{booktabs}       
\usepackage{amsfonts}       
\usepackage{nicefrac}       
\usepackage{microtype}      
\usepackage{graphicx}
\usepackage{amsmath,amsfonts,amsthm}
\usepackage{array}
\usepackage{subcaption}
\usepackage{bbm}
\usepackage{bm}
\usepackage{multicol}
\usepackage{lipsum}
\usepackage{wrapfig}
\usepackage[ruled]{algorithm2e}
\usepackage{arydshln}
\usepackage{multirow}
\usepackage{enumitem}
\usepackage{makecell}
\usepackage{caption}

\newtheorem{remark}{Remark}

\newtheorem{example}{Example}

\newcommand{\Ical}{\mathcal{I}}

\newcommand{\Lcal}{\mathcal{L}}

\newcommand{\Gcal}{\mathcal{G}}

\newcommand{\Ebb}{\mathbb{E}}
\newcommand{\Rbb}{\mathbb{R}}

\begin{document}

\title{Causal Structure Learning with Recommendation System}

\author{Shuyuan Xu$^\dagger$, Da Xu$^\ddagger$, Evren Korpeoglu$^\ddagger$, Sushant Kumar$^\ddagger$} 
\author{Stephen Guo$^\S$, Kannan Achan$^\ddagger$, Yongfeng Zhang$^\dagger$}
 \affiliation{
  \institution{$^\dagger$Rutgers University\qquad $^\ddagger$Walmart Labs \qquad $^\S$Indeed}
    }
 \email{shuyuan.xu@rutgers.edu, {da.xu,EKorpeoglu,Sushant.Kumar}@walmart.com}
 \email{sguo@indeed.com,kannan.achan@walmart.com,yongfeng.zhang@rutgers.edu}

\renewcommand{\shortauthors}{Shuyuan Xu et al.}

\begin{abstract}
A fundamental challenge of recommendation systems (RS) is understanding the causal dynamics underlying users' decision making.
Most existing literature addresses this problem by using causal structures inferred from domain knowledge.
However, there are numerous phenomenons where domain knowledge is insufficient, and the causal mechanisms must be learnt from the feedback data.
Discovering the causal mechanism from RS feedback data is both novel and challenging, since RS itself is a source of intervention that can influence both the users' exposure and their willingness to interact.
Also for this reason, most existing solutions become inappropriate since they require data collected free from any RS.

In this paper, we first formulate the underlying causal mechanism as a causal structural model and describe a general causal structure learning framework grounded in the real-world working mechanism of RS. 
The essence of our approach is to acknowledge the unknown nature of RS intervention.
We then derive the learning objective from our framework and propose an \emph{augmented Lagrangian solver} for efficient optimization.
We conduct both simulation and real-world experiments to demonstrate how our approach compares favorably to existing solutions, together with the empirical analysis from sensitivity and ablation studies. 

\end{abstract}


\keywords{Recommender Systems, Causal Discovery, Explainability, Graphical Model, Structural Equation, Unknown Intervention}




\maketitle

\section{Introduction}
\label{sec:introduction}
In recent years, there has been growing interest in understanding how the actions taken by a recommendation system (RS) can induce changes to the subsequent feedback data.
This type of causal reasoning is critical to the explainability, fairness, and transparency of RS. 
In contrast to machine learning who primarily focuses on data-driven problem solving, causal discovery investigates into the data generating mechanism and tries to understand how the observations are formed.
Therefore, depending on the question of interest, the fact that RS itself can interfere with the users' feedback can be both troublesome and useful.

For example, most machine learning model assumes that the collected data is generated by a static distribution. 
However, making a recommendation is likely to intervene with the user's decision making, thus changing the potential feedback \cite{xu2022intervention}. 
Also, notice that RS interventions are often systematic, meaning they are designed by developers in specific patterns rather than being purely random, which is very different from ordinary statistical fluctuations \cite{bottou2013counterfactual}. 
Therefore, when it comes to machine learning and evaluation, those interventions will cause various types of bias, and a great deal of literature has been devoted to addressing such issues \cite{chen2020bias}.
On the other hand, discovering causal relationships can benefit significantly from systematic interventions. 
The reason is that when an intervention takes place during the data-generation process, the systematic changes it causes provides an opportunity for us to track down the underlying cause-effect mechanisms.

To our knowledge, discovering causal mechanisms with RS has rarely been studied previously due to various challenges.
Most existing solutions cannot handle unknown interventions made by RS that are unrelated to causal discovery.
Consequently, the RS community relies primarily on causal mechanism inferred from domain knowledge \cite{zhang2021causal,xu2021causal}, but they often lack the coverage, versatility, and ability to explain many phenomenons of interest. 
With causal inference emerging as a key instrument for many RS studies and applications, it becomes imperative to learn the desired causal mechanism from RS feedback data.

However, unlike other scientific fields (such as clinical trials, etc.) where interventions are purposefully made to elucidate the causal mechanisms of interest \cite{pearl2009causality}, the interventions made by RS are primarily designed to increase the users' engagement and revenue. 
Worse yet, we may not even be able to tell whether a recommendation has indeed changed the users' decision making, which means the intervention from RS is of an unknown nature.
As a result, with the existing solutions, it is nearly impossible to identify the underlying causal mechanisms using RS feedback data alone.

An important observation that makes causal discovery possible, even with unknown intervention, is that an effect given its causes remains invariant to changes in the mechanism that generates the causes \cite{peters2017elements}.
It implies that while the recommendation made by RS can interfere with what causes a user to give the feedback, the causal mechanism behind the user's decision making is unaltered regardless of the interference. The opposite statement is not true, that the occurrence of a cause given the effect will not be invariant under outside interference. As we discuss later, this critical asymmetry can help us identify the cause-effect relationship in the RS feedback data.

Rather than making unrealistic assumptions to consolidate the unknown interventions of RS, we propose a novel modelling technique through a mixture of competing mechanisms. 
The high-level intuition is similar to that of the classical mixture of distributions \cite{reynolds2009gaussian}, but we make two significant progress:
\begin{enumerate}[leftmargin=*,topsep=0pt]
    \item the expectations are now taken with respect to an \textbf{expert} -- which is a stochastic indicator function -- that judges the winner of the competing mechanisms, namely, the recommendation mechanisms and the causal mechanism;
    \item rather than using the traditional expectation-maximization (EM) optimization which is difficult to adapt to the mixture mechanism setting, we propose using the more advanced \textbf{Gumbel reparameterization} approach \cite{Jang17categorical} to address the gradient-over-expectation challenge.
\end{enumerate}
We present the underlying causal mechanism as a structural causal model (SCM), which consists of a set of structural equations and the associated causal graph, representing by a directed acyclic graph (DAG).
We also leverage the recent advances in learning DAG \cite{zheng2018dags}, and apply their continuous DAG constraint to our mixture-of-mechanism learning object.
We then integrate the above components and techniques with the augmented Lagrangian solver \cite{hestenes1969multiplier} that scales easily to causal mechanisms with hundreds of variables, whereas most existing solutions handle only dozens of variables. 

In Section \ref{sec:prelim}, we first develop a comprehensive view regarding how feedback data is generated under the intervention of RS and the causal mechanism of interest. We also provide the necessary background and relevant work in this section.
Then in Section \ref{sec:mechanism}, we further expand on the challenges and solutions of unknown RS interventions, and present the likelihood function (of users' decision making) as a mixture of competing mechanisms.
We present the complete causal structure learning procedure in Section \ref{sec:learning}, including the optimization algorithm.

We examine the proposed causal structure learning approach through both large-scale real-world RS datasets and simulation studies. 
In addition to examining the learnt causal structures, we also experiment on how causal structure learning can lead directly to improved recommendations. 
In the simulation, we reveal the accuracy of the proposed approach in recovering the ground truth causal mechanisms compared with the existing causal discovery solutions.
We summarize our contributions as follow:
\begin{itemize}[leftmargin=*,topsep=1pt]
    \item we describe a comprehensive causal structure learning framework under unknown RS intervention;
    \item we propose a principled causal structure learning solution via the mixture of competing mechanisms, and develop an efficient optimization procedure using reparameterization and the augmented Lagrangian method;
    \item the proposed approach is examined thoroughly via both real-data experiments and simulation studies.
\end{itemize}

\vspace{-5pt}
\section{Preliminaries and Related Work}
\label{sec:prelim}
We briefly introduce the data generating mechanism of RS, and the background of causal structure learning. We also discuss how our work connects to a wide spectrum of recent literature in information retrieval, causal inference, and machine learning. 

\vspace{-5pt}
\subsection{Feedback generation under RS}
\label{sec:feedback}

Feedback data of RS are often generated as a result of interventions made by the RS.
This statement applies to a wide range of RS settings where users are exposed to specific contents (e.g. product, video, news, ads) made available by the historical or current RS powered by some underlying algorithm. 
We focus mainly on the more general implicit feedback setting where the users' response does not suggest the degree of relevance, but is simply an indicator of active interaction. 
It will be apparent that our development extends directly to explicit feedback.

\begin{figure}
    \centering
    \includegraphics[width=\linewidth]{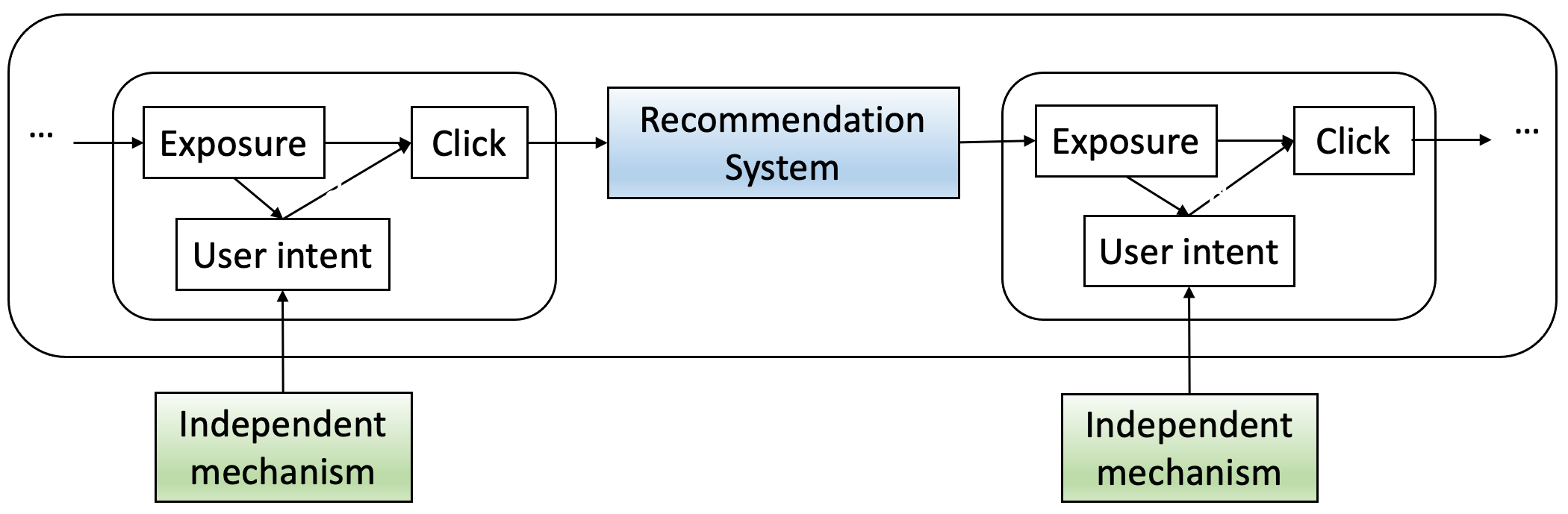}
    \vspace{-20pt}
    \caption{\small A simple illustration of how implicit feedback (click) is generated considering the impact of both the \emph{RS} and the \emph{independent (causal) mechanism} that underlies the user intent. 
    }
    \label{fig:recsys-mechanism}
    \vspace{-15pt}
\end{figure}

There is no doubt that users' decision making is a highly complex process resulting from the interaction of passive exposure and active human reasoning. 
We refer to this intrinsic interaction as \textbf{user intent}. 
We also use the notion of \textbf{mechanism} to refer to those modular, usable, and broadly applicable human intelligence for reasoning \cite{parascandolo2018learning}, and being \emph{independent} means they do not inform or influence each other.
For instance, the complementariness of \texttt{TV Cable} to \texttt{TV} is an independent mechanism since they are designed in such way by human intelligence.

We illustrate in Figure \ref{fig:recsys-mechanism} the interactive process of exposure, independent mechanism, user intent that eventually leads to the generation of user feedback.
We mention that the existence of an arrow in Figure \ref{fig:recsys-mechanism} only suggests the possibility to make an influence. 
A critical implication from Figure \ref{fig:recsys-mechanism} is that an interaction can be caused via multiple pathways. 
It means the exposure and the underlying independent mechanism may or may not have changed the user intent, and we are not able to tell the difference. 
As we discuss later, the essence of our approach is not assuming a pathway dominating the data generation process, which preserves the unknown nature of RS interventions.

\subsection{Characterizing causality and intervention}
\label{sec:causal-mechanism}

\begin{figure}[htb]
    \centering
    \includegraphics[width=\linewidth]{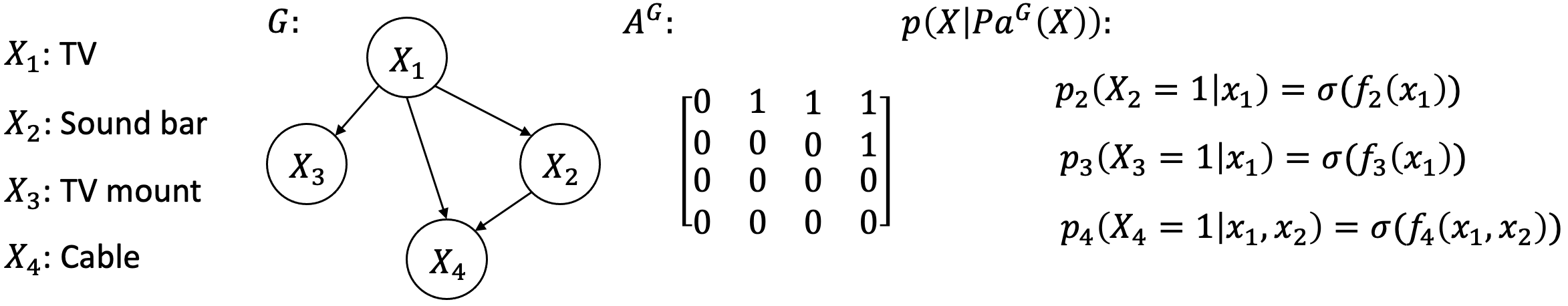}
    \vspace{-20pt}
    \caption{\small An exemplary structural causal model with four \emph{product types}. Causal structure learning aims to reveal both $A^{\Gcal}$ and $\{f_i\}_{i=2}^4$.}
    \label{fig:scm-example}
    \vspace{-0.3cm}
\end{figure}

Recent causal inference literature has established rigorously connection from independent mechanism to causality
\cite{peters2017elements,arjovsky2019invariant,parascandolo2018learning}.
The structural causal model, which consists of a joint distribution (can be further factorized into a set of structural equations) and the associated \emph{directed acyclic graph} (DAG), is often used to characterize the causal relationships among variables \cite{pearl2009causality}. 
In particular, the model is defined by a distribution $P_X$ over the random variables $X_1,\ldots,X_d$, and its factorization corresponds to the patterns of the DAG. 
Each node in the DAG corresponds to a random variables and each edge represents a direct causal relation. 
Given a DAG $\Gcal$ under which the joint distribution $P_X$ is \emph{Markovian}, it holds:
\begin{equation}
\label{eqn:CGM}
p_X(x_1,\ldots,x_d) = \prod_{j=1}^{d}p_j\Big(x_j \,\big|\, \text{Pa}^\Gcal(x_j)\Big),
\end{equation}
where $\text{Pa}^\Gcal(x_j)$ is the set of \emph{parent node} of $x_j$ according to $\Gcal$, and the conditionals $p_j, j=1,\ldots,d$ can be thought of as independent mechanisms that generate $x_j$ from its parents. In Figure \ref{fig:scm-example}, We show an example of a structural causal model for the \texttt{product type} relationships among some electronics. The (weighted) adjacency matrix of $\Gcal$ is denoted by $A^{\Gcal}\in\Rbb^{d\times d}$

Structural causal model provides us a framework to understand how the system responds to \emph{interventions} -- the systematic changes that are being made to the target distribution \cite{tian2013causal}. 
Following the example in Figure \ref{fig:scm-example}, if we fix $x_2=1$ throughout the experiment, then the new joint distribution will become: $\prod_{j\neq 2}p_j\Big(x_j \,\big|\, \text{Pa}^\Gcal(x_j)\Big)$.
For the feedback data collected under RS, the recommendations that were made are clearly interventions: they decided what were exposed to the user, thereby changing the feedback distribution systematically.
In contrast, irrelevant factors can only bring ordinary statistical fluctuations to the target distribution.

Making an intervention of the variable $x_j$ amounts to replacing its conditional distribution $p_j$ by a new $\tilde{p}_j$ in the joint distribution. 
In other words, the intervention modifies the joint distribution locally. 
There are two major types of interventions: the \emph{hard intervention} and the \emph{soft intervention}. 
Hard intervention means fixing a variable to the pre-determined value regardless of its causes, a scenario that may occur if we have perfect control of the experiment. 
For instance, we recommend \texttt{Television} to all customers regardless of everything else.
For this perspective, the influence from RS more resembles the \emph{soft} intervention, which is often milder such as \emph{changing the conditional probability of a variable given its causes}. The joint density after a soft intervention on $x_j$ is thus given by:
\[
p_{\text{int}}(x_1,\ldots,x_d) := \tilde{p}_j\Big(x_j \,\Big| \, \text{Pa}^\Gcal(x_j)\big) \prod_{q\neq j} p_q\Big(x_q \, \big| \, \text{Pa}^\Gcal(x_q)\Big),
\]
where we use $p_{\text{int}}$ to denote the \emph{intervened} distribution, $\tilde{p}_j$ to denote the \emph{soft intervened} conditional probability of variable $x_j$.

\subsection{Causal structure learning}
\label{sec:structure-learning}

Inferring the causal structure from empirical data has been a key research topic in many scientific disciplines \cite{pearl1995theory,spirtes2000causation}.
Using the fact that the conditional probability of an effect given its causes is invariant to interventions on the causes, the traditional discovery uses either \emph{constraint-based} or \emph{score-based} searching methods \cite{spirtes2000causation,peters2017elements}. However, they often scale poorly to the number of nodes, and the major blocker $\Gcal$ must be a DAG throughout the searching process.
Also, the search-based solutions cannot take advantage of the modern gradient-based optimization. 

The recent breakthrough by \citet{zheng2018dags} casts the search problem as a \emph{constraint continuous-optimization problem}. 
The key idea is that the acyclicity constraint on the weighted adjacency matrix of $\Gcal$ satisfies:
\begin{equation}
\label{eqn:DAG-constraint}
    \text{Tr}\Big(e^{A^{\Gcal}}\Big) - d = 0.
\end{equation}
Therefore, given the probability density function $p_{\theta}(\vec{X})$\footnote{We use $\vec{X}$ as a shorthand for $[X_1,\ldots,X_d]$} where $\theta$ is the structural parameter that induces the adjacency matrix $A^{\Gcal}_{\theta}$, the score-based method with the log-likelihood $\Lcal(\cdot)$ as objective function can be framed as:
\begin{equation}
\label{eqn:log-lik}
\max_{\theta} \Ebb_{\vec{X} \sim P_X} \Lcal\big(\theta \,|\, \vec{X}\big) \quad \text{s.t.} \quad \text{Tr}\Big(e^{A^{\Gcal}_{\theta}}\Big) - d = 0,
\end{equation}
which can be efficiently solved by an augmented Lagrangian procedure as proposed in \cite{zheng2018dags}.


\subsection{Relevance with existing work}
\label{sec:relevant-work}

\textbf{The causal view of RS.} Viewing RS as a causal system requires knowing the cause-effect relations among all variables involved \cite{bottou2013counterfactual,wang2020causal}. While prior knowledge of a system's working mechanism may lead to useful causal modelling, for instance, how bidding can cause the perturbation of price, such knowledge is often scarce.
Many phenomenons of interest require discovering causal relationships from data, or more precisely, combining the prior knowledge with data-driven learning \cite{vowels2021d}. 
Unfortunately, we find a lack of relevant idea in the RS literature.
The interventions from RS are known to cause various types of bias \cite{chen2020bias,schnabel2016recommendations,saito2020unbiased,bonner2018causal,wang2019doubly}, feedback loop effect \cite{mansoury2020feedback}, reachability issues \cite{dean2020recommendations}, and concerns about long term fairness \cite{creager2020causal}, etc. 
In general, RS intervention is often considered troublesome by those studies. 
In contrast, we believe RS intervention provides a valuable opportunity to discover the causal mechanisms that generate the feedback data.

\textbf{Causality-based recommendation.} 
Another key motivation for studying the causal structure behind RS is to leverage \emph{counterfactual reasoning} for recommendation \cite{mehrotra2018towards,joachims2016counterfactual,bottou2013counterfactual,xu2020adversarial}. 
The \emph{uplift modelling} is another rising topic that relies on on causality to optimize and evaluate recommendations \cite{sato2019action,sato2019uplift}.
Unfortunately, most studies in this direction have to rely on simple causal structures inferred from common sense, and the applications of counterfactual reasoning are primarily debiasing \cite{wei2021model,bonner2018causal}. 
Our causal learning techniques can provide insights into the above topics -- the more we understand the causal mechanisms behind the data, the better we can design machine learning and evaluation methods.

\textbf{Continuous causal structure learning.} 
Following the seminal work of \citet{zheng2018dags} which frames linear causal structure learning as a continuous optimization problem, a series of work extended their approach by incorporating neural networks to detect the more complex non-linear structures \cite{lachapelle2019gradient,yu2019dag,zheng2020learning}. 
More recently, continuous DAG optimizations are applied to learning causal structures in the presence of interventional data \cite{brouillard2020differentiable,squires2020permutation,ke2019learning}.
A major concern with this line of research is that they usually consider intervention-free data which is very rare in RS. 
A recent work \citet{wang2022sequential} combines sequential recommendation with continuous causal structure learning \cite{zheng2018dags}, but it assumes feedback data as intervention-free data. 
Most feedback data collected from information retrieval systems have went through RS interventions, which we will discuss further in Section \ref{sec:mechanism}.

\section{Characterizing (Unknown) Intervention Mechanism of RS}
\label{sec:mechanism}
For most implicit feedback setting, the collected data shows a trajectory of the user's interactions. 
When the user proceeds from the former item to the next, the interactive process and underlying decision-making generally correspond to the mechanism we introduced in Figure \ref{fig:recsys-mechanism}.
It involves the RS intervention (exposure), user intent, and independent mechanisms.

 \begin{figure}[htb]
    \centering
    \vspace{-10pt}
    \includegraphics[width=\linewidth]{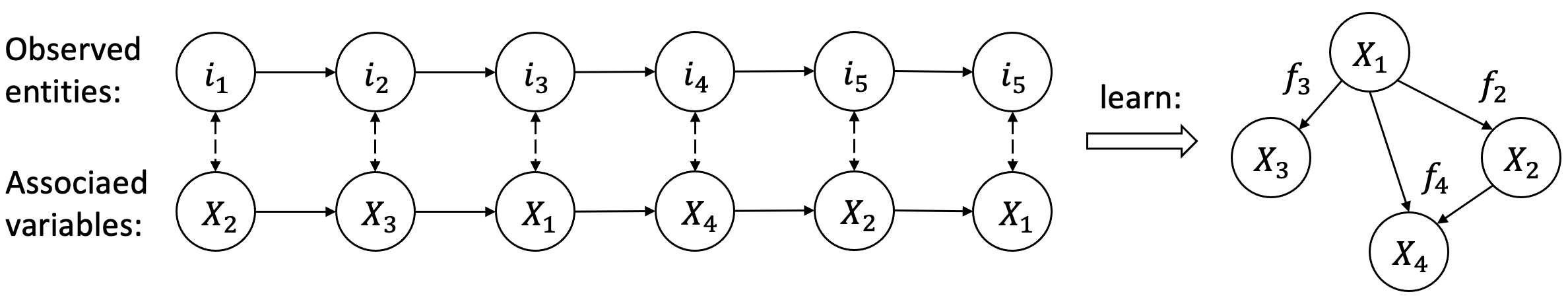}
    \vspace{-20pt}
    \caption{\small An illustration of learning causal structure from user feedback data.}
    \label{fig:illustration}
    \vspace{-0.3cm}
\end{figure}

We use $\Ical$ to denote the set of items such that $(i_1,\ldots,i_K), i_j\in\Ical$ for $j\in[K]$ is a trajectory of the user's interaction. 
We mention that the trajectory could be indexed by the user variables. But for the purpose of this paper, we focus on the causal structure among items (Figure \ref{fig:illustration} as an illustration).
We use the random variable $X_k(i) \in \{0,1\}$ to denote item $i$ possess the $k^{\text{th}}$ variable (or feature). For instance:, if $i\in\Ical$ is an article and $X_1,\ldots,X_d$ are the \emph{tags} (e.g. $X_1$ is for \texttt{basketball news}), then $X_1^{(i)}=1$ indicates that article $i$ is tagged with the \texttt{basketball news}. In this case, causal structure learning aims to reveal how reading articles of certain types can cause the reader become interested in another type of articles.
When the causal variables are continuous, we can simply use continuous distributions.

As we discussed previously, learning causal structure from user behavioral data requires characterizing how RS interferes with the user's decision making. 
The diagram in Figure \ref{fig:recsys-mechanism} provides a high-level illustration, and we need to rephrase the causal mechanism according to how RS works in practice. 
The is because many pioneer studies in causal inference have advocated that, the proposed causal structure should "model the reality" and "explain the objective world with generality" \cite{peters2017elements,pearl2018book}. 
Following this principle, we first examine a case study of a customer's online shopping journey at an e-commerce website -- an example that is general enough for understanding the intervention made by RS.

\begin{example}[Shopping online]
\label{example:shopping}
Suppose the customer's trajectory is given by: $i_{1:t}:= [i_1,\ldots,i_t]$. The next stages are:
\begin{enumerate}[leftmargin=*]
    \item The customer already has the next intended item $\tilde{i}$ in mind due to some underlying \emph{causal mechanisms}, driven by such as similarity, complementariness, and substitutability.
    \item The e-commerce webpage passes the data of $i_{1:t}$ to the RS, which instantly delivers a new set of recommendations to the webpage.
    \item Then one of the three following decisions will take place: \\
    \textbf{(a)}. $\tilde{i}$ is among of the recommendations and it becomes $i_{t+1}$;\\
    \textbf{(b)}. $\tilde{i}$ is not in the recommendations, but the user's mind is affected by the recommended items and select one of them as $i_{t+1}$;\\
    \textbf{(c)}. $\tilde{i}$ is not in the recommendations, and the user's mind is not changed. The user then searches for $\tilde{i}$ (which makes it $i_{t+1}$).
\end{enumerate}
\end{example}

The key takeaway from this example is that the RS intervention takes two steps. Making the exposure in \textbf{stage (2)} is the first step, and changing the user's mind as in \textbf{scenario (b)} is the second step.
Notice that user interacting with a recommended item does not necessarily a successful intervention, because the interaction may happen regardlessly (as in \textbf{scenario (a)}). 
In conclusion, a genuine \emph{intervention} from RS should have altered the customer's original intention which is driven by the underlying causal mechanisms. 

\begin{figure}[htb]
    \centering
    \vspace{-5pt}
    \includegraphics[width=\linewidth]{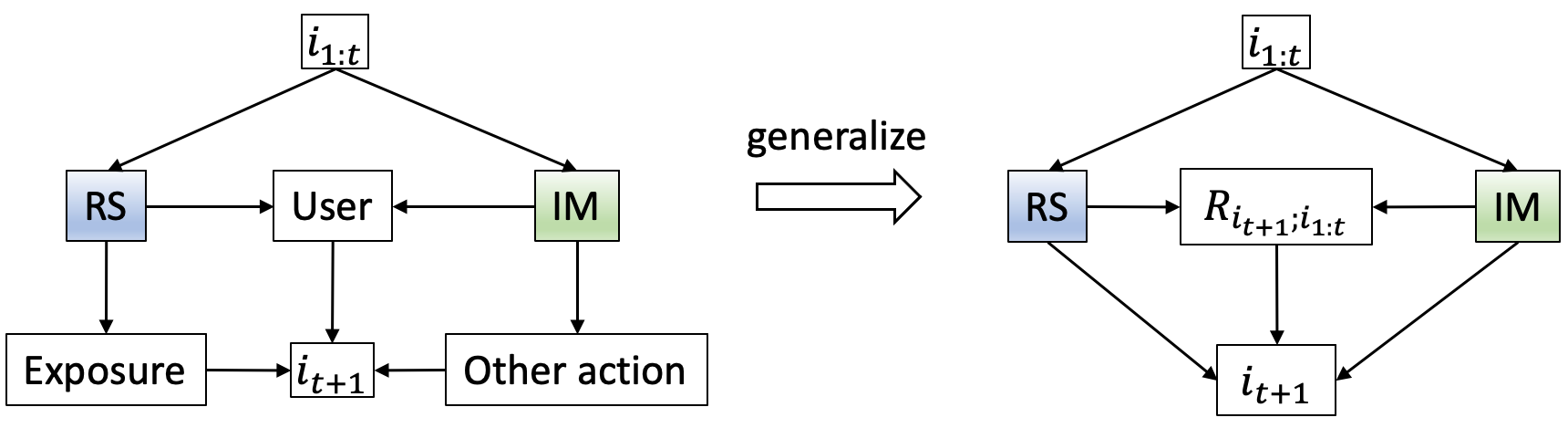}
    \vspace{-20pt}
    \caption{\small The causal diagram for the generating mechanism of RS feedback trajectory. The variable \emph{User} represents user intent, \emph{IM} represents the independent (causal) mechanism of interest, and $R_{i_{t+1},i_{1:t}}$ is an \textbf{indicator variable} suggesting whether RS or IM has made the true intervention. (Left): the causal diagram underlying the online shopping journey, where \emph{Other action} refers to such as the manual searching. (Right): the generalized structure that reduces to the pathways for: $i_{1:t}$, RS, IM, and $i_{t+1}$.}
    \label{fig:caual-graph}
    \vspace{-0.3cm}
\end{figure}

In the left panel of Figure \ref{fig:caual-graph}, we draw the \emph{causal diagram} (CD) that concludes our previous discussions. 
It is easy to check that the CD can be further generalized to a more compact version that depends entirely on $i_{1:t}$, the RS, and the causal mechanism (right panel of Figure \ref{fig:caual-graph}). 
This is made possible by bringing in the indicator variable $R_{i_{t+1},i_{1:t}}$, which indicates whether RS or IM has made a genuine intervention. 
One detail worth mentioning is that $R_{i_{t+1},i_{1:t}}$ is indexed by both $i_{t+1}$ and $i_{1:t}$. 
This is because the pathways from RS and IM to $i_{t+1}$ both involve $i_{1:t}$. 

In what follows, we use $X(i)$ to denote the item $i$'s variable\footnote{We assume for now that each item is associated with only one variable, however, our setting can extend directly to the multi-variate setting by further factorizing $P_X$.} such that $X(i_t) = X_j$ if and only if $X_j^{(i_t)}=1$. We use $\tilde{p}$ to denote the intervention mechanism of RS. We let $R_{i_{t+1},i_{1:t}}=1$ to denote a successful intervention from the RS and $R_{i_{t+1},i_{1:t}}=0$ for IM.
If $R_{i_{t+1};i_{1:t}}$ is observed, the conditional probability of $(i_1,\ldots,i_{t}) \to i_{t+1}$ is simply given by:
\begin{equation}
\label{eqn:conditional-pdf}
\begin{split}
& p\big(i_{t+1} \,|\, i_{1:t}, R_{i_{t+1},i_{1:t}}\big) \\
& = p\big(X(i_{t+1}) \,\big|\, \text{Pa}^{\Gcal}\big(X(i_{t+1})\big)\big)^{1-R_{i_{t+1};i_{1:t}}} \cdot \tilde{p}\big(i_{t+1} \,|\, i_{1:t}\big)^{R_{i_{t+1},i_{1:t}}},
\end{split}
\end{equation}
where the first term in the factorization represents the causal mechanism we try to discover, and the second term is how RS made the intervention. We can then parameterize the causal mechanism $p$ and RS mechanism $\tilde{p}$ and apply the maximum-likelihood estimation.

Unfortunately, we cannot infer directly from implicit feedback data whether RS has successfully intervened with the user's original decision making (\textbf{Scenario (b)} in Example \ref{example:shopping}). 
This particular issue has also been recognized by various RS studies, including \emph{positive-unlabelled} learning \cite{kato2018learning,bekker2020learning} and \emph{uplift modelling} \cite{sato2019action,sato2019uplift}. 
The solutions from those research often make
certain data assumptions that make inferring $R_{i_{t+1},i_{1:t}}$ easier, however, many of the assumptions cannot be substantiated \cite{xu2020adversarial}.

In our work, we directly treat the interference from RS as \textbf{unknown intervention}. It not only means the RS intervention mechanism $\tilde{p}$ is unknown, but more importantly, we acknowledge that $R_{i_{t+1},i_{1:t}}$ is unidentifiable as well. 
Nevertheless, we find treating RS intervention as unknown presents both challenge and opportunity for causal structure learning:
\begin{itemize}[leftmargin=*,topsep=0pt]
    \item on the one hand, we cannot optimize $p_X$ or $\tilde{p}$ directly since the log-likelihood function, which depends on $R_{i_{t+1},i_{1:t}}$, in unknown;
    \item on the other hand, we are free to consider $p_X$ and $\tilde{p}$ as two \emph{competing mechanisms}, where $R_{i_{t+1},i_{1:t}}$ acts as an \emph{expert} that decides which mechanism better explains the collected data under the mixture of distributions. 
\end{itemize}

Our innovation of using the mixture of \emph{competing mechanisms} is grounded in the machine learning literature \cite{cesa1997use,goyal2019recurrent}. 
The intuition is that the mechanism that best explains the data, in the presence of alternative explanations (mechanisms), is more likely to be the causal mechanism \cite{peters2017elements}. 
In other words, although we do not know in advance how the RS intervention worked, it still cause systematic changes to the data distribution, from which we can peek into the data-generating process.

\section{Learning Causal Structure from RS Feedback Data}
\label{sec:learning}
From Section \ref{sec:prelim}, a structural causal model consists of the causal graph described by it adjacency matrix $A^{\Gcal}$, and the structural equations with $f_j\big(\text{Pa}^{\Gcal}(X_j)\big) := p_j\big(X_j=1 \,|\, \text{Pa}^{\Gcal}(X_j)\big)$. By convention, we use $A_{k,j}$ to denote the  $(k,j)^{\text{th}}$ entry of $A$, and use $A_{j}$ to denote its $j^{\text{th}}$ row. 
For clarity, we directly use items to index the adjacency matrix or functions associated with their corresponding variables, e.g. $A^{\Gcal}_{i_t,i_k}$ and $f_{i_t}$. For instance, it holds: $A^{\Gcal}_{i_t,i_k} := A^{\Gcal}_{p,q}$ where $X_p = X(i_t)$ and $X_q = X(i_k)$. We use $\sigma(\cdot)$ to denote the sigmoid function.

\subsection{Likelihood for unknown RS intervention}
\label{sec:score-function}

To be consistent with the recent causal structure learning literature \cite{ke2019learning,brouillard2020differentiable}, we also treat each edge as sampled independently from a Bernoulli distribution with \emph{structural parameters} $\gamma_{k,j}$, such that $A^{\Gcal}_{k,j}\sim \text{Bernoulli}(\sigma(\gamma_{k,j}))$.
The structural equations $\{f_j\}_{j=1}^d$ are parameterized independently by linear or non-linear models such as the \emph{multilayer perceptron} (MLP). 
By using $\odot$ to denote the \emph{Hadamard product}, the expression of $\vec{X}\odot A^{\Gcal}_j$ gives exactly the set of parents for $X_j$. 
Therefore, we can rewrite $f_j$ using: $f_j\Big(\vec{X}\odot A^{\Gcal}_j\Big)$.

Recall that both the RS intervention mechanism (represented by $\tilde{p}(i_t \,|\, i_{i:t})$) and the success of the interventions (represented by $R_{i_{t+1},i_{1:t}}$) are unknown.
In some circumstances, $\tilde{p}$ can be directly given by the past recommendation policy if available.
Otherwise, we need to approximate the RS mechanism using some (sequential) recommendation algorithm such as \emph{GRU4Rec} \cite{hidasi2015session} or \emph{attention-based} model \cite{kang2018self}, depending on the problem instance.
When characterizing $R_{i_{t+1}, i_{1:t}}$, we have discussed its role in the previous section that it acts like an expert overseeing the two competing mechanisms. 
If the causal mechanism is unable to explain the data, which means none of $i_{1:t}$ is associated with the causal parents of $X(i_{t+1})$, then the expert should give credit to the RS intervention mechanism. 
Formally, we can characterize this competition between the two mechanisms as:
\begin{equation}
\label{eqn:R-formulation}
\begin{split}
    p\big(R_{i_{t+1}, i_{1:t}}=1\big) &= p\Big(A^{\Gcal}_{i_t,i_k}=0,\, \forall i_k \in i_{1:t}\Big) \\
    &= \prod_{i_k\in i_{1:t}} \big(1 - \sigma\big(\gamma_{i_t, i_k}\big) \big),
\end{split}
\end{equation}
which means RS intervention takes place when the causal mechanism is unable to explain the user's decision making.
For the sake of notation, we denote the last expression via the function $r(\Gamma)$ where $\Gamma := \big\{\gamma_{j,k}\big\}_{j,k=1}^d$ represents the set of structural parameters.
It follows that the distribution of $R_{i_{t+1}, i_{1:t}}$ is parameterized by $\Gamma$.
So far, our development consists of the following components:
\begin{itemize}[leftmargin=*,topsep=0pt]
    \item the \textbf{causal graph} that follows $A^{\Gcal}_{k,j}\sim \text{Bernoulli}(\gamma_{k,j})$, which we denote by the short hand $A^{\Gcal}\sim \sigma(\Gamma)$;
    \item the \textbf{structural equations} $f_j\Big(\vec{X}\odot A^{\Gcal}_j\Big) = p_j\big(X_j=1 \,|\, \text{Pa}^{\Gcal}(X_j)\big)$ parameterized by such as linear model or MLP;
    \item the \textbf{RS intervention mechanism} $g(i_{1:t+1}) := \tilde{p}(i_t \,|\, i_{i:t})$ parameterized by such as GRU4Rec;
    \item the \textbf{intervention indicator variable} $R_{i_{t+1}, i_{1:t}}$ that is sampled according to (\ref{eqn:R-formulation}), which we denote by the shorthand: $R\sim r(\Gamma)$.
\end{itemize}

Now we are ready to derive the likelihood (score) function for user's decision making process from $(i_1,\ldots,i_t) \to i_{t+1}$. In particular, we express the score function under unknown RS intervention as an expectation over the stochastic indicator function $R$:
\begin{equation}
\begin{split}
    & \Lcal\big(\vec{X} \,\big|\, \Gamma, \{f_j\}_{j=1}^d, g\big) \\
    & := \Ebb_{R\sim r(\Gamma)} \log p\big(i_{t+1} \,|\, i_{1:t}, R\big) \\ 
    & = \Ebb_{A^G\sim \sigma(\Gamma), R\sim r(\Gamma)} \log \Big[f_{i_{t+1}}(\vec{X}\odot A^{\Gcal}_{i_{t+1}})^{1-R} \cdot g(i_{1:t+1})^R  \Big].
\end{split}
\end{equation}
It follows that the score function is a mixture of the two competing mechanisms: the \emph{causal mechanism} and the \emph{RS mechanism}. 
The structural parameters $\Gamma$ has two roles in the score function: \textbf{1)}. generate the adjacency matrix $A^{\Gcal}$ which reveals the causal parents for each variable via $\vec{X}\odot A^{\Gcal}_{i}$; \textbf{2)}. inform the expert $R$ how likely the intervention was made by each competing mechanism. 
The parameters in the score function consists of: 
\begin{enumerate}
    \item the structural parameters $\Gamma=\{\gamma_{j,k}\}_{j,k=1}^d$ for the graph;
    \item the parameters for the structural equations $\big\{f_j(\cdot)\big\}_{j=1}^d$;
    \item the parameters for the RS mechanism $g(\cdot)$.
\end{enumerate}

Before we present the optimization algorithm for our proposed score function, we wish to discuss how the score function can adapt to a wide range of RS settings where the existing methods become special cases of our approach.

\begin{remark}[A note on model complexity]
\label{remark:complexity}
Some readers may have concerns for our model complexity due to the number of structural equations involved. We point out that the causal graph will be sparse as we will add regularizations on $\Gamma$ during training. Therefore, only a small number of causal structure equations will be truly active.
\end{remark}

\subsection{Optimization algorithm}
\label{sec:optimization}

We aim to develop an efficient optimization procedure that can handle large-scale causal structure learning. Most existing solutions are demonstrated on causal graphs with dozens of variable \cite{zheng2018dags,zheng2020learning,ke2019learning,brouillard2020differentiable}, which is considerably small compared with the settings in RS. For instance, modern social media, forum, or e-commerce platform can be hundreds or thousands of \texttt{tags} or \texttt{product types}.

Using the continuous DAG constraint introduced in Section \ref{sec:prelim}, the learning objective follows:
\begin{equation}
\label{eqn:constraint-objective}
    \underset{\Gamma,\, \{f_j\}_{j=1}^d,\, g}{\sup} \Ebb_{\vec{X}\sim P_X}\Lcal\big(\vec{X} \,\big|\, \gamma, \{f_j\}_{j=1}^d, g\big) \,\text{  s.t.  }\, \text{Tr}\Big(e^{\sigma(\Gamma)} - d \Big)=0.
\end{equation}

The objective function poses two challenges for optimization: 
\begin{enumerate}[leftmargin=*]
    \item the objective function presents a \textbf{constraint optimization} problem with non-linear constraint;
    \item the expectation w.r.t. $A^G\sim \sigma(\Gamma)$ and $R\sim r(\Gamma)$ in $\Lcal$ renders the objective function \textbf{intractable} (there is no explicit expression).
\end{enumerate}

To solve the first challenge, we adopt the \emph{augmented Lagrangian} procedure proposed in \cite{zheng2018dags}. The essence is to transform the constraint problem into a sequence of \emph{unconstrained problems} as:
\begin{equation}
\label{eqn:augmented-lagrangian}
\underset{\Gamma,\, \{f_j\}_{j=1}^d,\, g}{\sup} \Ebb_{\vec{X}\sim P_X}\Lcal\big(\vec{X} \,\big|\, \gamma, \{f_j\}_{j=1}^d, g\big) - \lambda^{(t)} h(\Gamma) - \frac{\mu^{(t)}}{2}h(\Gamma)^2,
\end{equation}
where we use $h(\Gamma):= \text{Tr}\Big(e^{\sigma(\Gamma)} - d \Big)$ to represent the penalty on violating the constraint, and use $\lambda^{(t)}$ and $\mu^{(t)}$ as the Lagrange multipliers for the $t^{\text{th}}$ unconstraint problem. For each unconstraint subproblem, we have two options for address the challenge of intractable expectation:
\begin{itemize}[leftmargin=*]
    \item apply the \emph{Monte Carlo estimator} with the log-trick, which is also known as the \emph{REINFORCE} estimator \cite{rezende2014stochastic};
    \item apply the "reparameterization trick" on Bernoulli sample by using such as the \emph{Straight-Through Gumbel-Softmax} estimator \cite{Jang17categorical}. 
\end{itemize}
We choose the later because it is known to have a lower variance and better computation complexity. 
In particular, the reparameterized sample of $s\sim \text{Bernoulli}(\sigma(\gamma))$ is given by:
\[
\mathbf{1}\big(\sigma(\gamma + l) \geq 0.5 \big) + \sigma(\gamma + l) - GB(\sigma(\gamma + l)),
\]
where $\mathbf{1}$ is the indicator function, $l$ is an independent sample from the standard \emph{Logistic distribution}, and $GB(\cdot)$ is a function such that $GB(u)=u$ and $\nabla_u GB(u) = 0$. In this way, the reparameterized sample still evaluates to the Bernoulli sample with probability given by $\sigma(\gamma)$, and the gradient w.r.t. $\gamma$ can be directly computed using the reparameterized sample.

Using the Gumbel reparameterization trick, we can efficiently compute the gradient of: $\nabla_{\Gamma}\Lcal\big(\vec{X} \,\big|\, \gamma, \{f_j\}_{j=1}^d, g\big)$, and apply the stochastic gradient descent algorithm such as \emph{RMSprop} to update the parameters. Note that the gradients w.r.t. to the parameters of $\{f_j\}_{j=1}^d$ and $g$ can be obtained in the same fashion. 

Suppose we find the approximate optimum $\Gamma^{(t)}$ for the $t^{\text{th}}$ unconstraint problem, following the augmented Larangian algorithm \cite{zheng2018dags}, the update rules for $\lambda$ and $\mu$ are:
\begin{equation}
\label{eqn:lagrangian-multipliers}
\begin{split}
    & \lambda^{(t+1)} = \lambda^{(t+1)} + \mu^{(t)}\cdot h(\Gamma^{(t)}); \\
    & \mu^{(t+1)} = \left\{ 
    \begin{aligned}
    \eta\cdot\mu^{(t)}, & & \text{if } h(\Gamma^{(t)}) > \delta\cdot h(\Gamma^{(t-1)}) \\
    \mu^{(t)} & & \text{otherwise},
    \end{aligned}
    \right.
\end{split}
\end{equation}
and we use $\eta=2$ and $\delta=0.9$ throughout our experiment.

Although the algorithm may not lead to the global optimum due to the non-convexity of the objective function, the algorithm should converge to a stationary point of (\ref{eqn:constraint-objective}). After obtaining the solutions $\Gcal^*$ and $\big\{f_j^*\big\}_{j=1}^d$, for example, we can use $f_j^*\Big(\vec{X}\odot A^{\Gcal^*}_j\Big)$ to infer the probability of observing $X_j=1$ given the other variables that represents the user's past interactions. The operation of $\odot$ acts like a \emph{masking} that maintains only the active causal pathways.

\section{Experiments and Results}
\label{sec:experiment}
We examine the proposed causal structure learning approach using both the real-data experiments and simulation study. 
We elaborate on the superior performance of using the discovered causal structures to make recommendations. All the reported results are averaged over \textbf{five} independent runs.

\subsection{Dataset, settings, and tasks}
\label{sec:dataset}

\subsubsection{{\textbf{Real-data experiments}}}

We conduct experiments on the \texttt{Electronics} department of the \emph{Amazon} \cite{he2016ups} and \emph{Walmart} \cite{xu2020knowledge} dataset. \footnote{We do not use Yahoo R3 and Coat datasets, which are benchmark datasets for counterfactual reasoning for recommendation, because they do not have product metadata thus not suitable for our causal structure learning task.}. 
The causal variable of interest is the items' \textbf{product type} (\textbf{PT}), given by such as \texttt{TV}, \texttt{Computer}, \texttt{Cell Phone}. 
In other words, we aim to discover the network of causalities behind the product types.
We specifically use the \texttt{Electronics} department because:
\begin{enumerate}[leftmargin=*]
    \item \texttt{Electronics} is the largest segment in the two datasets, and they both have rich feedback data;
    \item compared with others departments such as \texttt{books}, \texttt{movies}, and \texttt{clothes}, electronics products often exhibit stronger causal relationship as a result of their industrialized design, e.g. \texttt{TV}$\to$\texttt{Remote Control}, \texttt{XBox}$\to$\texttt{Handle}, \texttt{Phone}$\to$\texttt{Charger}.
\end{enumerate}

We now describe the details of the datasets and their processing.

\begin{itemize}[leftmargin=*]
    \item \textbf{Wmt Electronics:} The dataset consists of customers' same-session check baskets with the sequentially added products, and the product metadata includes the product type information. The feedback sequences are implicit by nature.
    \item \textbf{Amzn Electronics:} The dataset includes customers' rating sequences and the product metadata. The metadata includes concepts of products such as the \texttt{product type}. We convert ratings into the implicit feedback, where positive ratings as treated positive feedback by convention. 
\end{itemize}

The summary statistics for the datasets are provided in Table \ref{tab:datasets}. For preprocessing, we filter out infrequent product types with less than \emph{five} total occurrences.  

\begin{table}[hbt]
    \centering
    \resizebox{0.9\columnwidth}{!}{
    \begin{tabular}{ccccc}
    \toprule
        Dataset & \# Users & \# Items & \# PT & \# Interactions \\\midrule
        Wmt Electronics & 21,954  & 20,586 & 591 & 846,382\\
        Amzn Electronics & 28,386 & 14,471 & 874 & 660,159\\\bottomrule
    \end{tabular}
    }
    \caption{\small Summary statistics of real-world datasets}
    \label{tab:datasets}
    \vspace{-0.7cm}
\end{table}

\textbf{Real-data Tasks}.
The product type is a crucial concept in e-commerce that groups products with similar functionalities. 
Inferring the next \texttt{product type} that the customer may interact plays a critical role in the \emph{Homepage recommendations} for both \emph{Amazon.com} and \emph{Walmart.com}. 
Therefore, we treat it as our \textbf{recommendation task} where we use the users' past interactions to make recommendation sequentially.
With the proposed causal structure learning method, we can directly leverage the learnt conditional probabilities $p_j\big(X_j \,\big| \, \text{Pa}^{\Gcal}(X_j)\big),j=1,\ldots,d$ to rank the candidate product types. 
In particular, if product type $k$ is among $\text{Pa}^{\Gcal}(X_j)$ according to the learnt graph and it has been interacted by the user, then we have $x_k=1$. Otherwise, we have $x_k=0$.

We chronologically sort the feedback data and apply \textbf{leave-the-last-one-out} to split the training, validation and test data. Following the evaluation on top-$K$ recommendation task, we evaluate the models on \textbf{Hit@1}, \textbf{Hit@5}, \textbf{NDCG@5}, and the mean reciprocal rank (\textbf{MRR}). Since the product type is a segmentation variable, measuring the performance on the shorter length of recommendation list will be more indicative of the true performance. 
For evaluation, we apply \textbf{real-plus-N} \cite{said2014comparative} schema to calculate the value of each metric where we randomly sample 100 non-interacted product types as negative signal.

The causal graph is an important component for a structural causal model, thus \textbf{graph recovery} (i.e., how well the causal structure learning model can recover the ground-truth causal graph) is also a crucial task for causal structure learning model. Since we cannot access the ground truth causal graph on real data, we take several screenshots of the learnt graph which contain the product types most readers will find familiar. The more rigorous graph recovery will be conducted via simulation. 

\subsubsection{\textbf{Simulation study}}

The main purpose of simulation is to examine \textbf{graph recovery}, where we compare the learnt causal graph with the ground-truth causal mechanism that generates the synthetic data. 
When generating the synthetic feedback data, we directly follow the the RS working process depicted in Figure \ref{fig:recsys-mechanism}. 
We first simulate a RS given by the \emph{attention-based recommendation model} \cite{liu2018stamp} under random initialization. 
We specifically use \emph{different} RS models for simulation and causal structure learning (i.e. \emph{attention-based} versus \emph{GRU-based}) to avoid the potential bias. 
The ground-truth causal graph is give by a random DAG, where each edge is retained with a certain probability $p_{\text{keep}}$ while not violating the DAGness. 
We use the \textbf{structural Hamming distance} (\textbf{SHD}) as the evaluation metric to reveal the closeness between the learnt causal graph and the ground-truth graph \cite{brouillard2020differentiable}. In particular, SHD computes the number of edges that differ between two DAGs (either reversed, missing or superfluous).

When generating the feedback data, we assume there are \emph{10,000 users} each with \emph{15 interactions} total interactions. From the beginning to the end, the RS mechanism successfully intervene with each user under a probability of \emph{0.3}, which leads to the user randomly clicking one item from the recommended list of ten items. The total number of items equals to the number of nodes (causal variables) in the graph.
Otherwise, they interact with the item deemed most likely by the ground-truth causal mechanism, whose size we vary to create different settings (see Table \ref{tab:simulation}). The modelling, learning, and evaluation are the same as the real-data experiments.

\subsection{Baseline and model configurations}
\label{sec:baseline-config}

To provide a comprehensive comparison, we choose baseline models that are representative of: 1). collaborative filtering (CF) recommendation; 2). sequential recommendation; 3). causal structure learning. 
Since our approach and standard causal structure learning methods do not model users, for fair comparison, we consider only the item-base CF and sequential recommendation. 
\begin{itemize}[leftmargin=*]
    \item \textbf{GRU4Rec} \cite{hidasi2015session}. The classical session-based recommendation model which uses recurrent neural networks to capture the sequential patterns.
    
    \item \textbf{Item2vec} \cite{barkan2016item2vec}. A prevalent shallow embedding model inspired by word2vec \cite{mikolov2013distributed} that uses customers' sequential behavior data to embed items in a latent space for item-based CF.
    
    \item \textbf{NOTEARS} \cite{zheng2018dags}. The seminal causal graph learning model that firt frames the DAG requirement as a continuous constraint, and learn the causal structure based on observational data via continuous optimization. 
    
    \item \textbf{SDI} \cite{ke2019learning}. A recently proposed causal structural learning model that applies to the combination of observational and unknown interventional data, by alternatively learning the structural parameters and functional parameters.
    
\end{itemize}

We use the same training, validation and testing sets for all the models. \emph{GRU4Rec} adopts the \emph{Bayesian Personalized Ranking (BPR)} \cite{rendle2012bpr} framework for pair-wise learning. For each sequence $i_{1:t}$, we sample one product type not interacted clicked by the user as the negative target. 
For all the recommendation models, we fix the input embedding dimension as 64.
As for the other configuration of \emph{GRU4Rec}, we select the hidden dimension of the GRU from \{4,8,16,32,64\}, according to the validation performance.

In terms of the proposed approach (denoted by \textbf{CSL4RS}), we use a two-layer MLP to model each conditional probability, (i.e. the structural equations $\{f_j\}_{j=1}^d$ in (\ref{eqn:constraint-objective})), where the hidden dimension is selected from \{2,4,8,16,32\}. 
We also use \emph{GRU4Rec} to characterize the RS mechanism $g(\cdot)$ in our framework. It takes the \emph{5-most recent} interaction as the input sequence, and the GRU's hidden dimension is also selected from \{4,8,16,32,64\}.
We consider the following variants (\emph{ablated versions}) of our approach: \textbf{CSL4RS(Lin)} replaces the MLP by linear model;
\textbf{CSL4RS(-RS)} removes the RS intervention component; 
\textbf{CSL4RS(-CM)} removes the causal mechanism component.
We apply \emph{LeakyReLU} as activation function between layers for MLP. Additionally, we take learning rate as 0.001 and the $\ell_2$-regularization weight is 1e-6. 

For the \textbf{SDI} baseline, we chronologically split the user's data into two halves, where the first half is considered as observational data and the second half is considered as interventional data. We adopt the same model configurations and hyper-parameters as the published implementation of \citet{ke2019learning}. Finally, the model structure of \textbf{NOTEARS} is relatively simple so we keep the configurations as provided in its published implementation \cite{zheng2018dags}.

\begin{table}[htb]
    \centering
    \vspace{-5pt}
    \caption{The performances of the proposed approach and baseline methods on the recommendation tasks with \emph{Amzn} and \emph{Wmt} Electronics. The best results are highlighted in bold. In the parenthesis is the standard deviation.}
    \subcaption*{Amzn Electronics}
    \resizebox{1.03\columnwidth}{!}{
    \begin{tabular}{c|ccccc}
        & \emph{Item2vec} & \emph{GRU4Rec} & \emph{NOTEARS} & \emph{SDI} & \emph{CSL4RS} \\ \hline
         \textbf{Hit@1} & .1623(.02) & .1678(.01) & .1696(.01) & .1542(.01) & \textbf{.2607}(.02)\\
        \textbf{Hit@5} & .4552(.02) & .5149(.02) & .4142(.01) & .4540(.02) & \textbf{.5462}(.02) \\
        \textbf{NDCG@5} & .3121(.01) & .3615(.02) & .2957(.01) & .3065(.02) & \textbf{.3736}(.02) \\ 
        \textbf{MRR} & .3072(.02) & .3489(.01) & .2888(.01) & .3004(.01) & \textbf{.3607}(.02) \\ \hline 
    \end{tabular}}
    \subcaption*{Wmt Electronics}
    \resizebox{1.03\columnwidth}{!}{
    \begin{tabular}{c|ccccc}
        & \emph{Item2vec} & \emph{GRU4Rec} & \emph{NOTEARS} & \emph{SDI} & \emph{CSL4RS} \\ \hline
         \textbf{Hit@1} & .1928(.01) & .2454(.01) & .0956(.00) & .1689(.01) & \textbf{.2676}(.01) \\
        \textbf{Hit@5} & .4080(.02) & .4446(.02) & .2985(.01) & .3693(.02) & \textbf{.4864}(.02)  \\
        \textbf{NDCG@5} & .2779(.01) & .3195(.01) & .1866(.00) & .2415(.00) & \textbf{.3480}(.01) \\ 
        \textbf{MRR} & .2659(.01) & .2836(.01) & .2271(.01) & .2465(.01) & \textbf{.3014}(.01) \\ \hline 
    \end{tabular}
    }
    \label{tab:results}
    \vspace{-0.3cm}
\end{table}

\subsection{Performance analysis}
\label{sec:reco-performance}

\begin{figure}[htb]
    \centering
    \includegraphics[width=\linewidth]{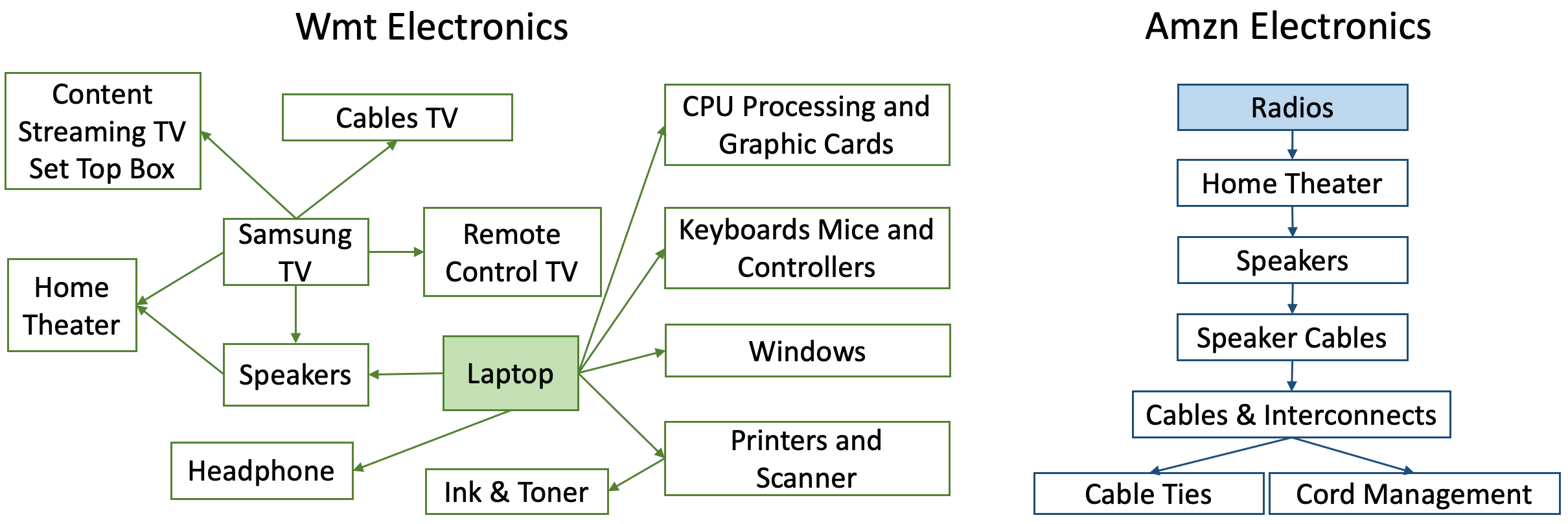}
    \vspace{-20pt}
    \caption{\small Graph examination on \emph{Amzn} and \emph{Wmt} Electronics. The learnt causal graph $\Gcal$ is constructed according to the optimized structural parameters $\gamma_{k,j}$, and locate the subgraphs whose parent nodes are \emph{Laptop} and \emph{Radios}. We particularly select \emph{Laptop} and \emph{Radios} as the parent nodes so readers can understand the quality of the learnt causal graph using common sense.}
    \label{fig:discovered-CGM}
    \vspace{-0.2cm}
\end{figure}

\textbf{ I) Compare with standard recommendation models}. From Table \ref{tab:results}, we observe that for both datasets, \emph{CSL4RS} significantly outperforms both \emph{Item2vec} and \emph{GRU4Rec} on all the metrics. 
This result suggests the following take away. 
Firstly, identifying causal patterns has a vast potential to improve recommendation performance -- although \emph{CSL4RS} makes recommendation based on individual MLPs who does not process sequential signal, it can still outperform GRU4Rec. 
Secondly, while standard recommendation models are known to detect associations in the data, the causation that drives the observed associations could be more useful.

\textbf{\ II) Compare with general causal structure learning models}. The less satisfying performances from \textbf{NOTEARS} and \textbf{SDI} -- which are the two important general-purpose causal structural learning models -- suggest that standard causal structure learning may not be suitable for RS feedback data without major modifications. 

\begin{figure}[htb]
    \centering
    \includegraphics[width=\linewidth]{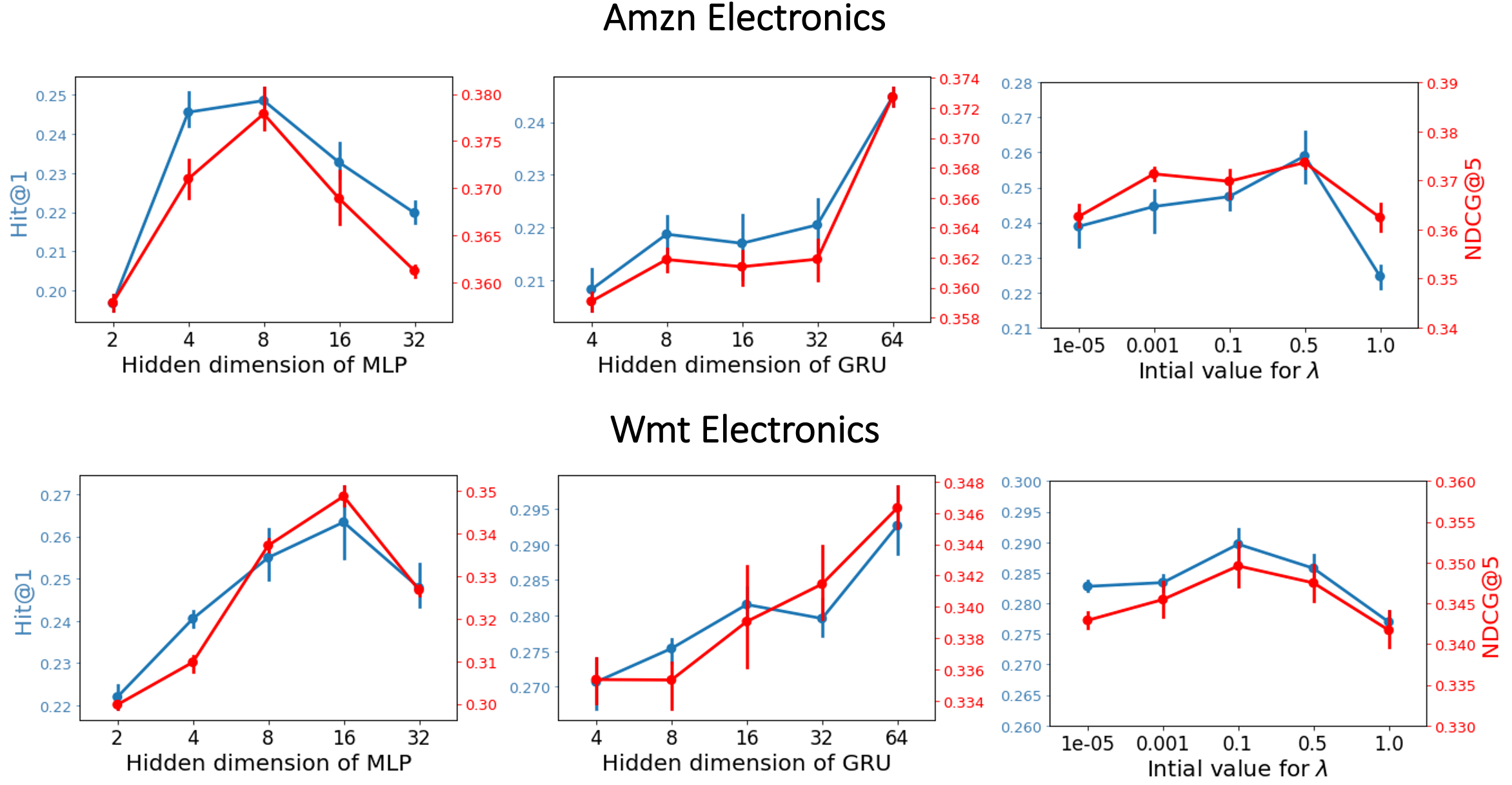}
    \vspace{-20pt}
    \caption{\small (\emph{Left panel}): sensitivity of the hidden dimension of MLP that model the causal structure equations; (\emph{Middle panel}): sensitivity of the hidden dimension of GRU that model the RS intervention mechanism; (\emph{Right panel}): sensitivity of the initial penalty parameter $\lambda^{(0)}$ for the augmented Lagrangian optimization.}
    \label{fig:sensitivity}
\end{figure}

\textbf{III) Ablation study}. We observe that using linear models for the structural equations $f_j(\cdot)$ (\emph{CSL4RS(Lin)}) downgrades the performance of our approach.
This is expected because linear model is less expressive than MLP, and it cannot capture the non-linear causal effects which are abundant in real-world data \cite{hoyer2008nonlinear}.
The fact that using MLP outperforms using linear model in our approach suggests our framework is indeed learning the underlying conditional probability models for $p_j\big(X_j \,\big|\, \text{Pa}^{\Gcal}(X_j))$. 
Otherwise, using MLP would not have made such big differences. 
The same conclusion can be drawn for the recommendation mechanism $g(\cdot)$, as Figure \ref{fig:sensitivity}b shows that using the more expressive \emph{GRU4Rec} model (by increasing its hidden dimension) consistently improves the performance. 

\textbf{IV) Sensitivity analysis}. Figure \ref{fig:sensitivity} suggests that up to a certain extend, increasing the expressively of MLP improves the performance. 
However, an over-large hidden dimension can lead to overfitting, which is a common issue for machine learning models. 
As for the initial optimization parameter $\lambda^{(0)}$ that penalizes violation to the DAG constraint, we observe an \emph{first-increase-then-decrease} pattern that is common to regularization parameters in machine learning.

\textbf{V) Real-data graph examination.} The ability to explain the data fundamentally differentiates causal structure learning from standard machine learning. 
Towards this goal, we examine the learnt causal graphs as shown in Figure \ref{fig:discovered-CGM}. 
An advantage of using the electronics datasets is that readers can easily judge how the learnt causal graph matches the common-sense causality of e-commerce products. 
While this result is not exhaustive, it readily shows the our appraoch is able to detect meaningful causal structures from RS feedback data. 
The rigorous evaluations are discussed below.

\begin{figure}[htb]
    \centering
    \includegraphics[width=\linewidth]{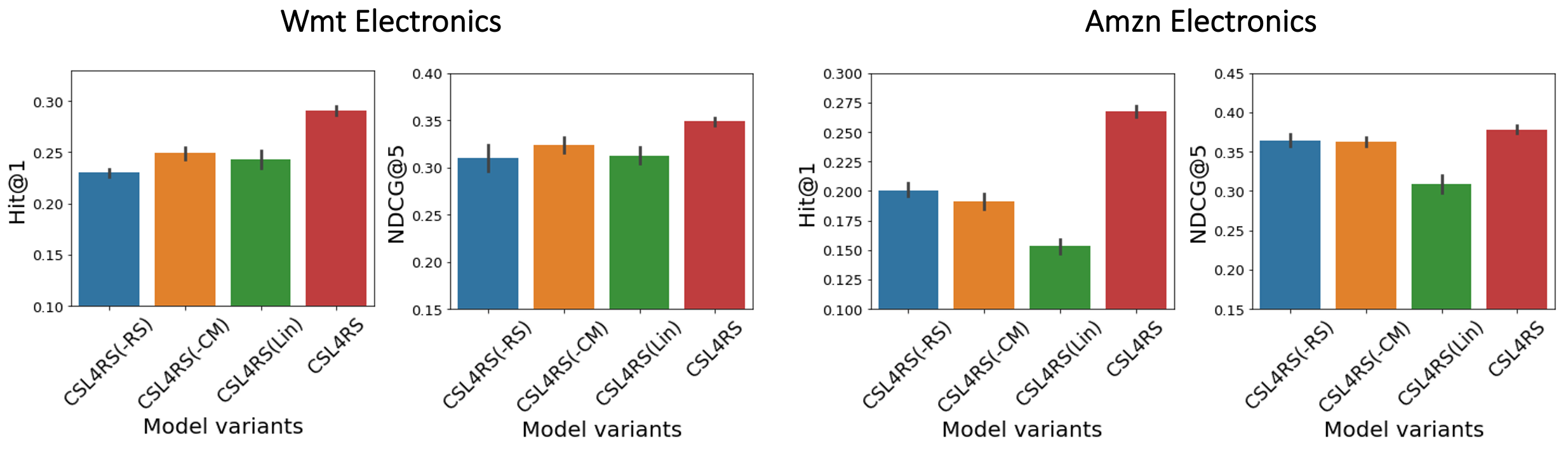}
    \vspace{-20pt}
    \caption{\small Ablation studies. Recall that CSL4RS(Lin) replaces MLP by linear model; CSL4RS(-RS) removes the RS intervention component; and CSL4RS(-CM) removes the causal mechanism component. }
    \label{fig:ablation}
    \vspace{-0.1cm}
\end{figure}

\textbf{VI) Simulation analysis.} 
From Table \ref{tab:simulation}, it is clear that \textbf{CSL4RS} significantly outperforms the other causal structure learning methods on the synthetic RS feedback data, particularly in the \textbf{graph recovery} task. 
Since we are comparing the learnt $\Gcal$ with the ground-truth causal graph, the results provide concrete evidence that our approach is truly capable of discovering the cause-effect relationships with RS feedback data. 
Also, \textbf{CSL4RS} gives the best performances in the recommendation task on our synthetic data. 
We mention that although our simulation setting is not the most sophisticated in the literature, it is sufficient to demonstrate \textbf{CSL4RS} as a capable and accurate causal discovery approach for RS.

\begin{table}[htb]
    \centering
    \resizebox{0.9\columnwidth}{!}{
    \begin{tabular}{c|cc|cc} \hline
        $p_{\text{keep}}=0.01$ & \multicolumn{2}{c}{50 nodes} & \multicolumn{2}{c}{100 nodes} \\ \hline
         & \textbf{NDCG@5} & \textbf{SHD} & \textbf{NDCG@5} & \textbf{SHD} \\ \hline
        \emph{NOTEARS} & .3462(.03) & 97.3(.2) & .3222(.04) & 198.4(.4) \\ 
        \emph{SDI} & .1997(.02) & 135.6(.8) & .1764(.01) & 496.8(.9)  \\
        \emph{CSL4RS(Lin)} & .2567(.05) & 73.3(.4) & .2469(.04) & 282.5(.4)  \\ 
        \emph{CSL4RS(-RS)} & .2623(.03) & 61.2(.4) & .3014(.04) & 352.6(.4)  \\ 
        \emph{CSL4RS} & \textbf{.4055}(.02) & \textbf{52.6}(.2) & \textbf{.3734}(.02) & \textbf{158.1}(.3) \\ \hline
    \end{tabular}}
    \caption{\small Graph examination results on the synthetic feedback data. Smaller \emph{SHD} indicates a better graph recovery. We keep the ground-truth graph relative sparse to mimic real-world causal mechanisms.}
    \label{tab:simulation}
    \vspace{-0.5cm}
\end{table}

\section{Conclusion}

We study in this paper the novel problem of learning causal structure with RS. The solution we proposed is grounded in the real-world working mechanism of RS and applies to a wide range of RS settings. 
More importantly, our approach handles the \emph{unknown nature of RS interventions} in a principled fashion, and the proposed optimization procedure scales easily to hundreds of causal variables. 

The real-data and simulation experiments both demonstrate that our approach not only dominates the causal discovery task, but also outperforms standard RS models in the recommendation tasks.
We believe causation can lead to major innovations in the way scientists perceive recommendations.
By establishing the solution that detects causal mechanisms from RS feedback data, our work can open the door to many future research directions in such as decision-making, debiasing, content understanding, and promoting explaniablity and fairness of RS.

\bibliographystyle{ACM-Reference-Format}
\bibliography{references}


\end{document}